\documentclass[runningheads]{llncs}
\usepackage[T1]{fontenc}
\usepackage{graphicx,verbatim}

\usepackage{amsmath}
\usepackage{amssymb}
\usepackage{booktabs}

\usepackage{multirow}

\usepackage{makecell}
\usepackage[hidelinks]{hyperref}

\begin{document}
\title{Seamless Whole Slide Label-Free Virtual Staining}
\author{Dou Hoon Kwark\inst{1} \and
Kianoush Falahkheirkhah\inst{1} \and
Ji-Hun Oh\inst{1} \and
Shirui Luo\inst{2} \and
Volodymyr Kindratenko\inst{1,2}\thanks{Contributed equally} \and
Rohit Bhargava\inst{1}$^\star$}

\authorrunning{D. H. Kwark et al.}
\institute{University of Illinois Urbana-Champaign, Urbana, IL, USA \and
National Center for Supercomputing Applications, Urbana, IL, USA \\
\email{dkwark2@illinois.edu}}
  
\maketitle           
\begin{abstract}
Label-free virtual staining offers a compelling, non-destruct\-ive alternative to standard histopathology; however, its clinical adoption is hindered by the computational bottlenecks inherent to processing gigapixel Whole Slide Images (WSIs). Current deep learning approaches require patch-based inference to avoid memory constraints, which disrupts global tissue continuity and introduces tiling artifacts—displaying visible seams and color shifts. To address this, we introduce the Consistency Memory Bank (COMB), a novel label-free virtual staining framework that enforces spatial and channel consistency across tiles without memory bottlenecks. COMB decouples context storage from computation, utilizing a dynamic retrieval mechanism to fetch feature representations from adjacent tiles. This enables a retrieval-based context integration strategy that adopts local padding to resolve spatial discontinuities and neighbor-aware channel attention to stabilize statistical drift. Further optimized with a sliding window schedule to ensure minimal memory overhead, our method demonstrates superior performance over state-of-the-art baselines, achieving significant improvements in both perceptual fidelity and tiling consistency, while suggesting its downstream utility in tumor segmentation. Code is available at \url{https://github.com/dou0000/COMB}.


\end{abstract}
\section{Introduction}
Pathology diagnosis relies on physical staining to make tissue structures visible under a microscope. While effective, this standard workflow is labor-intensive, inconsistent due to chemical variations, and consumes tissue samples required for downstream molecular assays. Label-free imaging \cite{Shaked:NatPhotonics:2023:LabelFree} offers a compelling alternative by visualizing tissue structures using their inherent signals, eliminating the need for destructive chemical dyeing process. The advent of virtual staining—using deep learning to translate raw data into histological images—offers a powerful way to optimize pathology workflows across diverse imaging modalities \cite{Rivenson:NBE:2019:AFVirtualStain,Schnell:PNAS:2020:IR_OH,Park:SciAdv:2025:Rapid,Chen:BOE:2021:Ultraviolet,Liu:SciAdv:2024:Virtual}

However, clinical adoption faces a major computational hurdle: memory constraints prevent the single holistic inference of gigapixel Whole-Slide Images (WSIs). To that end, recent models have been developed to process the image in independent patches. This fragmentation, however, disrupts global tissue continuity, introducing \textit{tiling artifacts}—spatial and photometric discontinuities that artificially cleave biological structures—which confound the visual continuity required for accurate pathological review.

Prior solutions for these tiling artifacts, widely adopted in virtual H\&E-to-IHC staining, rely on Instance Normalization (IN) \cite{Ulyanov:Arxiv:2016:InstanceNorm} augmented with global \cite{Chen:AAAI:2022:TIN} or adjacent statistics \cite{Ho:ECCV:2022:KIN,Ho:ECCV:2024:DenseNorm}. While IN works well for standard RGB-to-RGB tasks, it can potentially underperform in label-free generation (e.g., spectral signal-to-RGB). Since it normalizes statistics based on the local statistics (in contrast to Batch Normalization (BN), which relies on learned global statistics), IN can sometimes lead to unstable predictions, particularly in label-free source modalities where signal distributions can be sparse and heterogeneous. Consequently, BN \cite{Ioffe:ICML:2015:BatchNormalization} remains the standard for label-free virtual staining to ensure generation quality and stability \cite{Li:LSA:2021:BiopsyFreeSkin,Soker:Bioeng:2025:SpectralVirtualStain}.

Despite the stability of BN, mechanisms to enforce channel and spatial consistency across patches with respect to BN remain underdeveloped. The recent CC-WSI-Net \cite{Liu:ISBI:2025:CCWSINet} attempted to address this by building upon the VSGD-Net architecture \cite{Liu:WACV:2023:VSGDNet}—a strong generator integrating ResNet encoders and attention mechanisms—and adding global histogram matching to guide consistency and pixel-adjacency loss. However, this approach has limitations: global constraints often suppress local details \cite{Ho:ECCV:2024:DenseNorm}, and simple pixel-adjacency losses are ineffective when the input domain (e.g., spectral data) lacks direct visual correlation with the output (H\&E). Therefore, there is currently no single BN-based framework capable of delivering seamless, artifact-free generation for label-free virtual staining.

In this work, we introduce the \textbf{CO}nsistency \textbf{M}emory \textbf{B}ank (COMB), a novel framework designed to address the limitations of patch-based inference in label-free whole-slide virtual staining. COMB dynamically aligns adjacent patches to ensure visual continuity. Specifically, our contributions are as follows: (i) We present a novel seamless whole-slide generation framework tailored specifically for label-free virtual staining. (ii) We introduce a retrieval-based context integration strategy that integrates local padding \cite{Abdellatif:arXiv:2023:LocalPadding} and neighbor-aware Convolutional Block Attention Module to enforce spatial and channel consistency with minimal memory overhead. (iii) We validate that our method achieves superior performance on virtual staining quality and tiling artifacts compared to state-of-the-art baselines, with demonstrated utility in downstream tumor segmentation.

\begin{figure}[ht]
  \centering
  \includegraphics[width=0.7\textwidth,height=0.6\textheight,keepaspectratio]{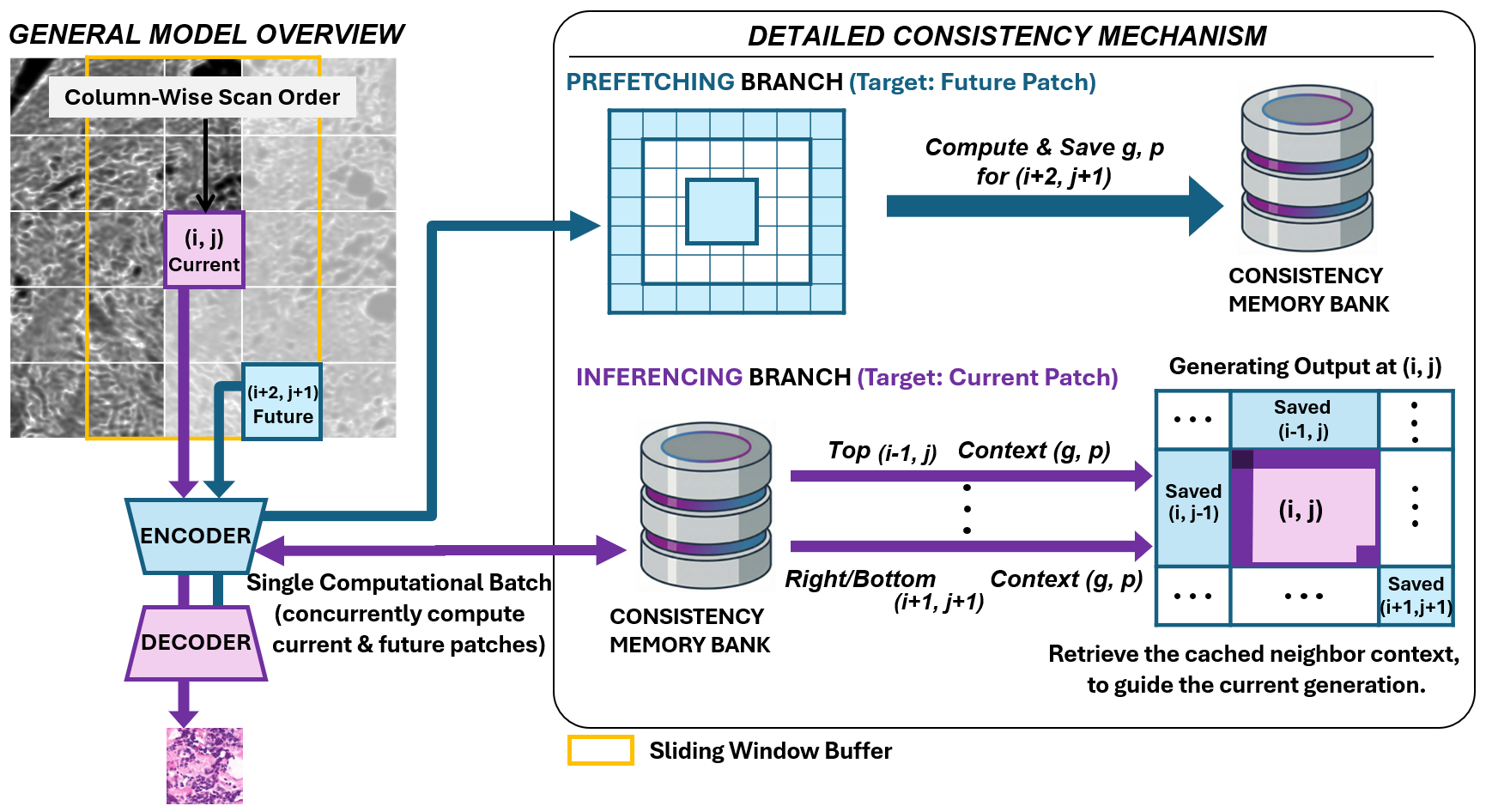}
  \caption{\textbf{COMB Architecture.} A sliding-window memory bank efficiently manages global context. Described for the context radius $R=1$ }
  \label{fig:model_fig_1}
\end{figure}

\section{Method}
\subsubsection{Problem Formulation.}
Let $x \in \mathbb{R}^{C_\text{in} \times H \times W}$ be a multi-spectral input and $y \in \mathbb{R}^{3 \times H \times W}$ be the corresponding H\&E slide. Our goal is to learn a generator $G_\theta$ such that $\hat{y}=G_\theta(x)$ is locally faithful (preserving cellular morphology and staining) and globally seamless. To handle gigapixel WSIs, we partition $x$ into a grid of smaller tiles $\{x_{i,j}\}_{i,j}$ of size $T \times T$ (e.g. $T=256$ or $512$). 

While such partition enables gigapixel WSI processing, it introduces two critical artifacts: (i) spatial discontinuities (seams) due to truncated receptive fields at tile borders, and (ii) channel-wise drift caused by the discrepancy between global training statistics and local inference windows.

\subsection{Consistency Memory Bank (COMB)}
To resolve these, building upon VSGD-Net \cite{Liu:WACV:2023:VSGDNet}, we introduce COMB, a novel label-free virtual staining framework that decouples context storage from computation, enabling efficient retrieval of consistent spatial and channel information to ensure seamless generation. Specifically, as shown in the model overview in Fig.~\ref{fig:model_fig_1}, features from neighboring tiles are extracted by the encoder and decoder and stored in COMB. This memory bank is dynamically maintained using a sliding-window mechanism, retaining only the necessary features to avoid redundancy. During inference on a given patch, COMB retrieves the relevant neighboring features as contextual guidance, allowing the current generation to leverage consistent spatial and channel information and ensuring spatial coherence across tiles.

\begin{figure}[ht]
  \centering
  \includegraphics[width=0.8\textwidth,height=0.8\textheight,keepaspectratio]{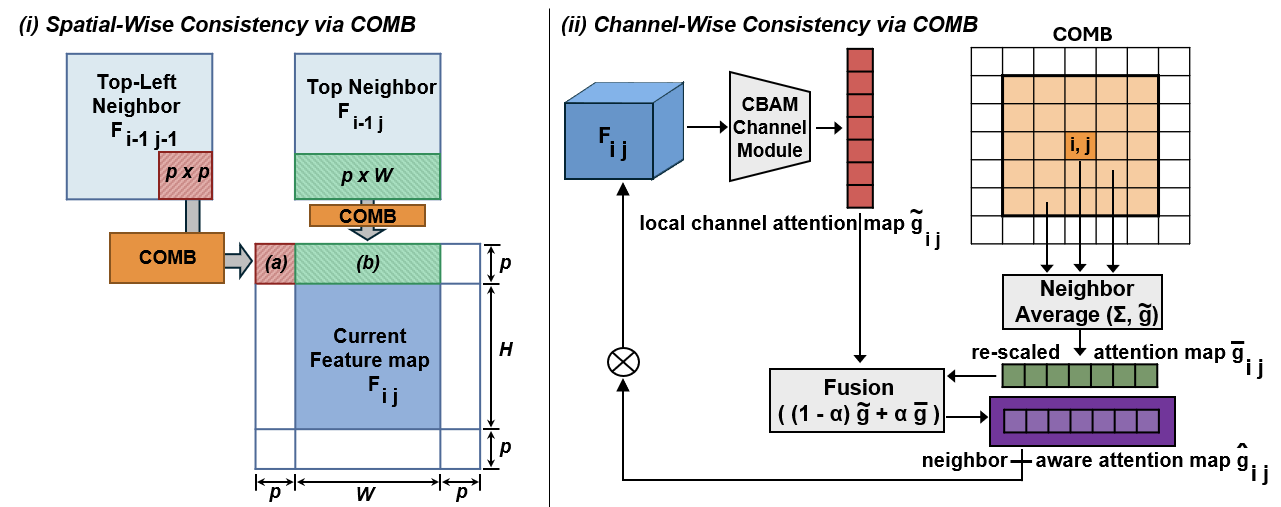}
  \caption{Retrieved features enforce (i) spatial consistency via local padding and (ii) channel consistency via neighbor-aware CBAM.}
  \label{fig:model_fig_2}
\end{figure}

\noindent \textbf{Spatial-Wise Consistency.}
Standard frameworks typically rely on explicit padding (e.g., reflection) or implicit zero-padding (e.g., the default zero-padding in standard convolution) to maintain feature map dimensions. However, these introduce synthetic boundary signals that are disconnected from adjacent patches. To address this, we implement local padding~\cite{Abdellatif:arXiv:2023:LocalPadding} via our proposed retrieval mechanism. As shown in Fig.~\ref{fig:model_fig_2}(i), we explicitly pad the margins of size $p$ by querying COMB for valid context—specifically, retrieving the boundary features from the full spatial neighborhood, utilizing stored history for processed tiles and pre-fetched representations for future context. This ensures the subsequent convolution operates on real, continuous tissue signals rather than synthetic artifacts.

\noindent \textbf{Channel-Wise Consistency (Neighbor-Aware Attention).}
While spatial padding ensures morphological continuity, it does not correct the statistical drift caused by the discrepancy between the learned global training statistics and the current local inference windows. To mitigate this, we propose a neighbor-aware Convolutional Block Attention Module (CBAM)~\cite{Woo:ECCV:2018:CBAM} integrated within the skip connections. Standard CBAM enhances virtual staining by preserving local details during encoder-to-decoder transfer \cite{Liu:WACV:2023:VSGDNet}. However, it calculates attention weights based solely on the isolated patch, making it blind to regional inconsistencies. Consequently, it often amplifies local anomalies (e.g., stain variations) rather than correcting them. To address this, we reformulate the module to perform dynamic feature recalibration (Fig.~\ref{fig:model_fig_2}(ii)). By aligning the attention mechanism with the retrieved neighbor context, we enforce regional consistency while retaining authentic tissue signals. First, we compute the standard local channel attention map $\tilde{\mathbf{g}}_{ij}$ using the isolated tile features $F_{ij}$ via Global Average Pooling (GAP) and Global Max Pooling (GMP):
\begin{equation}
\tilde{\mathbf{g}}_{ij} = \sigma\left(\operatorname{MLP}\left(\operatorname{GAP}(F_{i j})\right) + \operatorname{MLP}\left(\operatorname{GMP}(F_{i j})\right)\right),
\end{equation}
where $\sigma$ denotes the sigmoid function. To harmonize the local representation with its surroundings, we compute a regional attention map $\bar{\mathbf{g}}_{ij}$ from the neighbor set $\mathcal{N}_{ij}$ (tiles within the context radius $R$ of $(i,j)$) and use it to stabilize the final attention weights $\hat{\mathbf{g}}_{ij}$:
\begin{equation}
  \bar{\mathbf{g}}_{ij} = \frac{1}{|\mathcal{N}_{ij}|}\sum_{(p,q)\in\mathcal{N}_{ij}} \tilde{\mathbf{g}}_{pq},
  \qquad 
  \hat{\mathbf{g}}_{ij} = (1 - \alpha) \tilde{\mathbf{g}}_{ij} + \alpha \bar{\mathbf{g}}_{ij},
\end{equation}
where $\alpha$ is a hyperparameter. This fusion ensures that if the local signal $\tilde{\mathbf{g}}_{ij}$ is weak or anomalous, the model leverages the stable neighborhood map $\bar{\mathbf{g}}_{ij}$ to produce consistent inference.

\noindent \textbf{Concurrent Prefetching and Inference Scheduling.}
A naive implementation of COMB requires a two-pass strategy: first caching the global context, then performing inference. However, this effectively doubles the computational cost. To overcome this, we adopt a concurrent prefetching schedule~\cite{Ho:ECCV:2024:DenseNorm}, enabling COMB to populate contexts in a single continuous scan. Processing in column-major order, we construct a compound mini-batch at each step containing the current inference tile at $(i,j)$ and a look-ahead prefetch tile at $(i + R + 1, j + R)$, as illustrated in Fig.~\ref{fig:model_fig_1}. In this setup, the prefetch tile computes and caches features into the bank, while the inference tile retrieves these pre-populated contexts to generate the final output $\hat{y}_{ij}$. This ensures the necessary neighbor contexts are available in memory just-in-time for the inference step. However, this concurrent scheduling necessitates initializing the entire global context map to accommodate the sparse updates, which is significantly expensive for gigapixel WSIs. To address this, we further optimize the memory footprint by implementing COMB as a sliding window buffer rather than a global registry. We maintain only the active columns required by the context radius $R$, evicting stale columns as the inference front progresses. This reduces the memory complexity from $O(\frac{H}{T} \cdot \frac{W}{T})$ to $O(R \cdot \frac{H}{T})$ in the number of cached tiles (and thus in cached feature memory), making high-resolution WSI processing feasible on standard consumer hardware.

\begin{figure}[t]
  \centering
  \includegraphics[width=\textwidth]{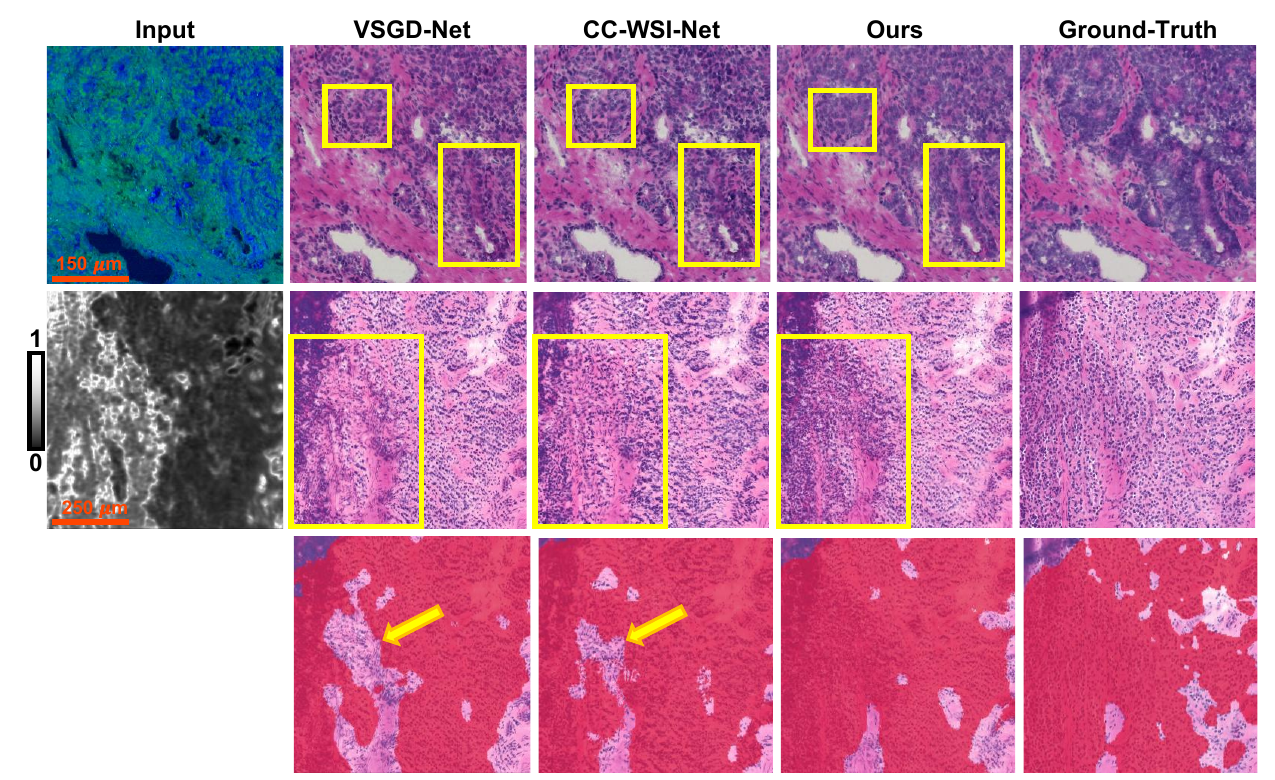}
  \caption{Visual comparison on SRS (top) and IR (middle) datasets. The third row represents segmentation prediction on the second row.}
  \label{fig:viz}
\end{figure}

\section{Experiments}

\noindent \textbf{Dataset.} We validated our method on a cohort of frozen prostate surgical resection samples (collected and IRB approved at Mayo Clinic), comprising two label-free modalities: multispectral mid-infrared (IR) and Stimulated Raman Scattering (SRS). The IR data was acquired using Laser Scanning Confocal Microscopy (LSCM) \cite{Yeh:NatComm:2023:Infrared} across ten discrete wavenumbers, while the SRS \cite{Falahkheirkhah:CRC:2023:Accelerating} data consists of five spectral channels; both sets were specifically selected to capture molecular signatures of proteins, nucleic acids, lipids, and carbohydrates. All spectral images were spatially co-registered with their corresponding H\&E stained counterparts and resized to a unified resolution of 0.5 $\mu$m/pixel. For preprocessing, we discarded patches where the non-tissue area exceeded 80\%. The final datasets comprise approximately 60,000 IR-HE and 16,500 SRS-HE paired $512 \times 512$ patches. For evaluation, we employed a 3-fold cross-validation strategy, ensuring strict patient-level separation between folds.

\noindent \textbf{Implementation Details.} 
We adopt the VSGD-Net \cite{Liu:WACV:2023:VSGDNet} generator, modified to accept 10-channel IR or 5-channel SRS inputs, and a multi-scale PatchGAN \cite{Isola:CVPR:2017:Pix2Pix} discriminator, optimized using the LSGAN objective \cite{Mao:ICCV:2017:LSGAN}. For perceptual loss, we replace the standard VGG-19 \cite{Johnson:ECCV:2016:PerceptualLoss,Simonyan:ICLR:2015:VGG} with a ConvNeXt-Tiny backbone \cite{Liu:CVPR:2022:ConvNeXt} fine-tuned on histopathology data to better capture domain-specific textures. Training is performed using Adam \cite{Kingma:ICLR:2015:Adam} ($\beta_1=0.5, \beta_2=0.999$) with a learning rate of $2\times10^{-4}$ and batch size of 8 for 200k iterations on an NVIDIA H200. For COMB, we set the memory bank parameters to $\alpha = 0.3$ and $R=2$.

\begin{table}[t]
  \centering
  \caption{Quantitative comparison on SRS-HE and IR-HE datasets. \textbf{Bold} indicates best, \underline{underline} indicates second best. Results reported as Mean\textsubscript{Std Dev}.}
  \label{tab:metrics}
  
  \small
  
  \begin{tabular}{clcccc}
    \toprule
    \textbf{Dataset} & \textbf{Metric} & \textbf{Pix2Pix} & \textbf{VSGD-Net} & \textbf{CC-WSI-Net} & \textbf{Ours} \\
    \midrule
    \multirow{4}{*}{\textbf{SRS-HE}} 
    & SSIM $\uparrow$
    & 0.4467\textsubscript{.031}
    & 0.4809\textsubscript{.032}
    & \underline{0.4830}\textsubscript{.028}
    & \textbf{0.4995}\textsubscript{.043} \\
    \addlinespace[0.2em]
    
    & LPIPS $\downarrow$
    & 0.4413\textsubscript{.024}
    & 0.4208\textsubscript{.012}
    & \underline{0.4141}\textsubscript{.016}
    & \textbf{0.4112}\textsubscript{.014} \\
    \addlinespace[0.2em]
    
    & FSG $\downarrow$ \textsubscript{(GT: 0.0752)}
    & 0.3960\textsubscript{.131}
    & \underline{0.2500}\textsubscript{.065}
    & 0.2978\textsubscript{.096}
    & \textbf{0.1054}\textsubscript{.026} \\
    \addlinespace[0.2em]
    
    & TexTile $\uparrow$ \textsubscript{(GT: 0.8453)}
    & 0.3047\textsubscript{.070}
    & \underline{0.4058}\textsubscript{.008}
    & 0.3743\textsubscript{.078}
    & \textbf{0.7320}\textsubscript{.066} \\
    \midrule
    \multirow{4}{*}{\textbf{IR-HE}} 
    & SSIM $\uparrow$
    & 0.4846\textsubscript{.033}
    & \underline{0.4902}\textsubscript{.030}
    & 0.4860\textsubscript{.026}
    & \textbf{0.4960}\textsubscript{.020} \\
    \addlinespace[0.2em]
    
    & LPIPS $\downarrow$
    & 0.4788\textsubscript{.004}
    & \underline{0.4764}\textsubscript{.008}
    & 0.4772\textsubscript{.008}
    & \textbf{0.4743}\textsubscript{.009} \\
    \addlinespace[0.2em]
    
    & FSG $\downarrow$ \textsubscript{(GT: 0.1009)}
    & 0.3556\textsubscript{.036}
    & 0.3321\textsubscript{.035}
    & \underline{0.2669}\textsubscript{.010}
    & \textbf{0.1080}\textsubscript{.011} \\
    \addlinespace[0.2em]
    
    & TexTile $\uparrow$ \textsubscript{(GT: 0.8491)}
    & 0.2500\textsubscript{.010}
    & 0.2896\textsubscript{.010}
    & \underline{0.3698}\textsubscript{.003}
    & \textbf{0.7654}\textsubscript{.009} \\
    \bottomrule
  \end{tabular}
\end{table}

\noindent \textbf{Experimental Setup.} We benchmark our framework against three representative baselines: Pix2Pix~\cite{Isola:CVPR:2017:Pix2Pix} as the foundational generative model; VSGD-Net~\cite{Liu:WACV:2023:VSGDNet}, a robust patch-based architecture; and CC-WSI-Net~\cite{Liu:ISBI:2025:CCWSINet}, a state-of-the-art virtual-staining framework for global consistency. To ensure spectral stability, we utilize Batch Normalization across all models, with the exception of CC-WSI-Net which retains its native hybrid architecture (BN encoder, IN decoder). During evaluation, WSIs are processed into $512 \times 512$ patches with a stride of 496 (16-pixel overlap). We then construct $2\times2$ stitched montages to center tile boundaries in the field of view. We evaluate generation quality using SSIM~\cite{Wang:TIP:2004:SSIM} and LPIPS~\cite{Zhang:CVPR:2018:LPIPS}. In addition, we quantify stitching consistency using Focused Sobel Gradient (FSG)~\cite{Armstrong:BMVC:2025:STAIN}, which detects gradient discontinuities at interfaces, and TexTile~\cite{RodriguezPardo:CVPR:2024:TexTile}, a learned metric assessing intrinsic seamlessness.

\subsection{Results}
\label{qualitative}
\noindent \textbf{Virtual Staining Performance.}
We first assess the perceptual fidelity and seamlessness of the generated H\&E images on both IR and SRS modalities. As shown in Table~\ref{tab:metrics}, our framework consistently outperforms baselines across all metrics. Notably, on the seamlessness metrics (FSG and TexTile), our method shows significant improvement over other methods by a substantial margin, closely approaching the metric numbers of the Ground Truth. These quantitative results are visually confirmed in Fig.~\ref{fig:viz}. In the SRS-HE translation (top-row), baselines exhibit sharp visible seams in the indicated bounding box, which leads to structural discontinuities and geometric misalignments with corresponding regions in the GT. On the other hand, in the IR-HE translation (second-row), heterogeneous signal intensities in the indicated region lead baseline generations to compromise morphological definition and obscure local tissue textures compared to our method. This validates that our retrieval-based context integration effectively resolves tiling artifacts without compromising spectral-to-morphological translation fidelity in patch-based inference.

\begin{table}[t]
\centering
\caption{Segmentation performance. \textbf{Bold}: Best, \underline{Underline}: Second Best (exclu\-ding GT). Results reported as Mean\textsubscript{Std Dev}.}
\label{tab:seg}

\small

\begin{tabular}{lcccc|c}
\toprule
\textbf{Metric} & \textbf{Pix2Pix} & \textbf{VSGD-Net} & \textbf{CC-WSI-Net} & \textbf{Ours} & \textbf{GT} \\
\midrule
Dice $\uparrow$        
& 77.91\textsubscript{7.20}  
& 79.43\textsubscript{3.35}  
& \underline{80.64}\textsubscript{2.90}  
& \textbf{80.97}\textsubscript{4.04}  
& 81.25\textsubscript{2.73} \\

IoU $\uparrow$          
& 66.33\textsubscript{8.67}  
& 67.33\textsubscript{4.20}  
& \underline{69.05}\textsubscript{3.67}  
& \textbf{69.50}\textsubscript{5.55}  
& 70.00\textsubscript{3.54} \\

Sens. $\uparrow$       
& 78.45\textsubscript{11.6} 
& 78.13\textsubscript{2.38}  
& \textbf{84.57}\textsubscript{4.68}  
& \underline{80.83}\textsubscript{9.91}  
& 84.63\textsubscript{3.52} \\

Spec. $\uparrow$       
& 78.60\textsubscript{11.2} 
& \textbf{82.08}\textsubscript{8.59}  
& 74.92\textsubscript{10.3} 
& \underline{81.15}\textsubscript{10.9} 
& 75.12\textsubscript{6.54} \\

HD95 $\downarrow$      
& \underline{241.37}\textsubscript{29.0}
& 245.35\textsubscript{48.7}
& 252.25\textsubscript{47.1}
& \textbf{234.50}\textsubscript{29.2}
& 224.45\textsubscript{30.9} \\
\bottomrule
\end{tabular}
\end{table}

\noindent \textbf{Downstream Segmentation.}
To verify clinical utility, we evaluated tumor segmentation performance on the generated images (refer to rows 2-3 in Fig.~\ref{fig:viz} and Table~\ref{tab:seg}), utilizing a UNet++~\cite{Zhou:DLMIA:2018:UNet++} trained on real H\&E images for each fold. Consistent with our qualitative analysis, the specific artifacts identified in baseline images directly confound the segmentation model: the structural discontinuities (seams) act as artificial boundaries that fragment the prediction masks, while the uncorrected statistical drift (triggered by local signal anomalies) distorts the tumor morphology, leading to misclassification. In contrast, our seamless output restores structural integrity, enabling precise delineation of tumor regions. This is important, as it influences not only the estimation of tumor size but also the characterization of infiltrative growth patterns reflected at the tumor boundary. Inaccurate representation of these features may introduce variability in risk stratification and could potentially affect assessments of cancer aggressiveness. Quantitatively, while yielding moderate improvements in Dice and IoU, our method achieves a superior balance between Sensitivity and Specificity, avoiding the skewed performance profiles observed in competing methods. Crucially, we significantly outperform all baselines in the $95\%$ Hausdorff Distance (HD95). As HD95 is highly sensitive to boundary outliers, this improvement confirms that our method effectively addresses the artifact-induced false positives that compromise automated analysis. 

\begin{figure}[t]
  \centering
  \includegraphics[width=1\textwidth,keepaspectratio]{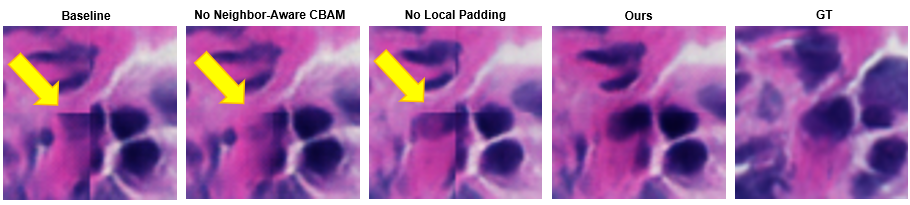}
  \caption{\textbf{Qualitative ablation study.} Component-wise ablation study shows th\-at ours effectively integrates both components for seamless, high-fidelity generation.}
  \label{fig:abl_viz}
\end{figure}

\noindent \textbf{Ablation \& Efficiency.} 
We validate our architectural contributions in Fig.~\ref{fig:abl_viz} and Table~\ref{tab:inference_time_memory}. As shown in Fig.~\ref{fig:abl_viz}, removing the neighbor-aware CBAM forces the model to rely on individual tiles. While local padding preserves cross-patch continuity, this lack of regional context triggers channel-wise drift, leading to washed-out generations and particularly degrading the reconstruction of centrally located nuclei. Conversely, while removing local padding yields better shape reconstruction for the isolated nuclei, it immediately reintroduces boundary-localized spatial seams due to truncated receptive fields. Thus, the full COMB framework is essential for achieving both seamlessness and high-fidelity generation. Furthermore, Table~\ref{tab:inference_time_memory} demonstrates the critical impact of our scheduling strategies. While concurrent scheduling significantly accelerates inference compared to the latency-heavy two-pass approach, it still shows a prohibitive memory cost. Our sliding window optimization resolves this bottleneck, reducing the memory footprint to a level close to the baseline with only a moderate increase in latency, thereby making gigapixel WSI processing feasible on standard consumer hardware.

\begin{table}
  \centering
  \caption{Inference performance on an NVIDIA RTX 2080 Ti ($30720 \times 28160$ WSI).}
  \label{tab:inference_time_memory}
  
  \footnotesize 
  \begin{tabular}{lccc|c} 
    \toprule
    \textbf{Metric} & \textbf{Two-pass} & \textbf{Concurrent} & \textbf{Sliding window} & \textbf{Baseline} \\
    \midrule
    Time [s] $|$ Mem [GB] & 302.4 $|$ 10.64 & \textbf{202.6} $|$ 10.64 & 262.8 $|$ \textbf{4.38} & 63.9 $|$ 3.50 \\
    \bottomrule
  \end{tabular}
\end{table}

\section{Conclusion and Future Work}
We presented the Consistency Memory Bank (COMB), a novel label-free virtual staining framework bridging memory constraints and global continuity in gigapixel whole-slide processing. Notably, COMB is the first fully Batch Norm-based approach to successfully resolve tiling artifacts, preserving the generation stability essential for label-free translation. By decoupling context retrieval from computation, COMB effectively mitigates these artifacts via retrieval-based local padding and neighbor-aware CBAM. Extensive evaluations on IR and SRS datasets demonstrate that our method surpasses state-of-the-art baselines in perceptual quality and generation seamlessness, and preserving this structural continuity translates to clinical utility, significantly reducing artifact-induced errors in downstream tumor segmentation. Despite these advances, our current implementation applies local padding uniformly across the network. A promising avenue for future work involves investigating the trade-off between optimal patch dimensions and selective context retrieval. By restricting local padding to only essential architectural layers, we anticipate driving further memory optimization without compromising the seamlessness of the whole-slide generation.

\subsubsection*{Acknowledgements.} 

This work used the Delta and DeltaAI systems at the National Center for Supercomputing Applications (NCSA) through allocation MED240056 from the Advanced Cyberinfrastructure Coordination Ecosystem: Services \& Support (ACCESS) program, and was supported in part by the Illinois Computes project (supported by the University of Illinois Urbana-Champaign and the University of Illinois System) and in part by the National Institutes of Health (NIH) under grant number R01CA260830.

\subsubsection*{Disclosure of Interests.} 
The authors have no competing interests to declare.

\end{document}